\documentclass[pra,aps,showpacs,twocolumn,amsmath,amssymb,floatfix]{revtex4}

\usepackage{hyperref}

\usepackage{graphicx}

\usepackage{amssymb,amsmath}
\usepackage{subfigure}
\usepackage{bbm}
\usepackage[usenames]{color}
\begin{document}

\title{Quark-Gluon Plasma at RHIC and the LHC:  Perfect Fluid too Perfect?}

\author{James L. Nagle$^{1,2}$, Ian G. Bearden$^{2}$, William A. Zajc$^{3}$}

\affiliation{$^1$Department of Physics, University of Colorado, Boulder, Colorado 80305, U.S.A.\\
$^2$Niels Bohr Institute, Discovery Center, University of Copenhagen, Copenhagen, Denmark\\
$^3$Department of Physics, Columbia University, New York City, New York 10027, U.S.A.}
\date{\today}

\begin{abstract}
Relativistic heavy ion collisions have reached energies that enable the
creation of a novel state of matter termed the quark-gluon plasma.
Many observables point to a picture
of the medium as rapidly equilibrating and expanding as a nearly
inviscid fluid.  In this article, we explore the evolution of
experimental flow observables as a function of collision energy and
attempt to reconcile the observed similarities across a broad energy regime
in terms of the initial conditions and viscous hydrodynamics.
If the initial spatial anisotropies are very similar for all collision
energies from 39 GeV to 2.76 TeV, we find that viscous hydrodynamics
might be consistent with 
the level of agreement for $v_2$ of unidentified hadrons as a function
of $p_T$.  However, we predict a strong collision energy dependence for the proton
$v_{2}(p_{T})$.
The results presented in this paper highlight the need
for more systematic studies and a {\em re-evaluation} of previously stated
sensitivities to the early time dynamics and properties of the medium.
\end{abstract}


\maketitle

\section{Introduction}\label{sec:introduction}
A physics case has been made that collisions at the Relativistic Heavy Ion Collider (RHIC) produce a
strongly-coupled system that evolves as a nearly perfect fluid
~\cite{Adcox:2004mh}; that is, the medium has a value of shear viscosity to
entropy density $\eta/s$ near a conjectured minimum bound~\cite{Son:2007vk}.
This categorization of the quark-gluon plasma as a nearly perfect
fluid has opened up interesting connections to other strongly-coupled
systems in nature~\cite{Johnson:29,thomas:34,jacak:39}.
Many features of the RHIC experimental data are well-described
by theoretical calculations using the assumption of
strong coupling. 
The most compelling of these results are  fully relativistic viscous hydrodynamic calculations 
that describe the  momentum anisotropy
patterns measured by experiment.  
The initial overlap geometry in these
nuclear collisions has a significant eccentricity for non-zero impact
parameters.  Both the general initial anisotropy and the event-by-event 
variations due to fluctuations in the nucleon positions
can be described
in terms of various moments $\epsilon_{2},
\epsilon_{3}, \epsilon_{4}, ...$~\cite{Alver:2010gr}.  If the created
medium were a non-interacting gas of particles,  these spatial
anisotropies would have no mechanism to translate into the momentum
distributions of partons and eventually final state hadrons. However,
in the limit of very strong interactions between the constituents 
(i.e. very short mean free paths), 
one expects substantial momentum anisotropies that might be
describable via modeling based on viscous hydrodynamics.  These momentum
anisotropies are often described in terms of measured Fourier moments
of the azimuthal distribution of particles -- $v_{2}, v_{3}, v_{4}, ...$~\cite{Ollitrault:1992bk}.
Extensive measurements of the even moments of these observables
by RHIC experiments indicate a strongly flowing fluid medium. 
Experimental efforts are now underway to study the odd Fourier flow coefficients
that result from fluctuations, motivated by the important observation of Refs.~\cite{Alver:2010gr,Sorensen:2010zq}.

\section{Energy Dependence of Perfection}

\begin{figure}[hb!]
\centering
\includegraphics[width=0.5\textwidth]{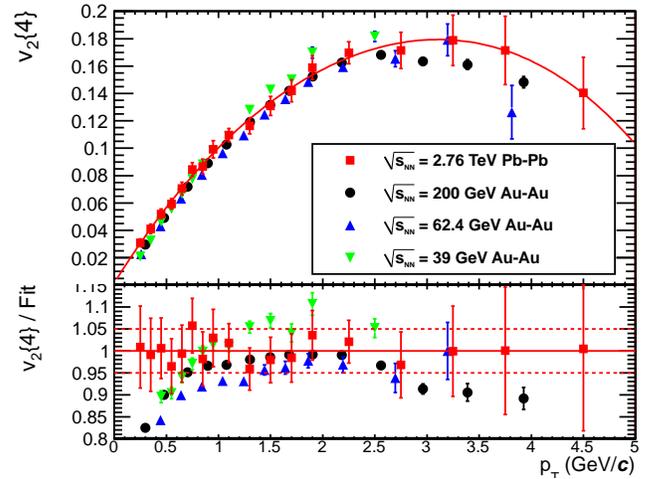}
\caption{Experimentally measured $v_{2}\{4\}$ as a function of transverse momentum $p_T$ for unidentified
  charged hadrons in the 20-30\% centrality selection.  Published results from the ALICE experiment at
  2.76 TeV~\cite{Aamodt:2010pa} and the STAR experiment at 200 GeV~\cite{Abelev:2008ed} are shown, along with
  preliminary results from lower energies~\cite{Kumar:2011de}.  The lower panel shows all
  data sets divided by a common fourth-order polynomial fit to the
  ALICE data points.
\label{fig_data_compare}}
\end{figure}

\begin{figure*}[ht]
\centering
\includegraphics[width=1.0\textwidth]{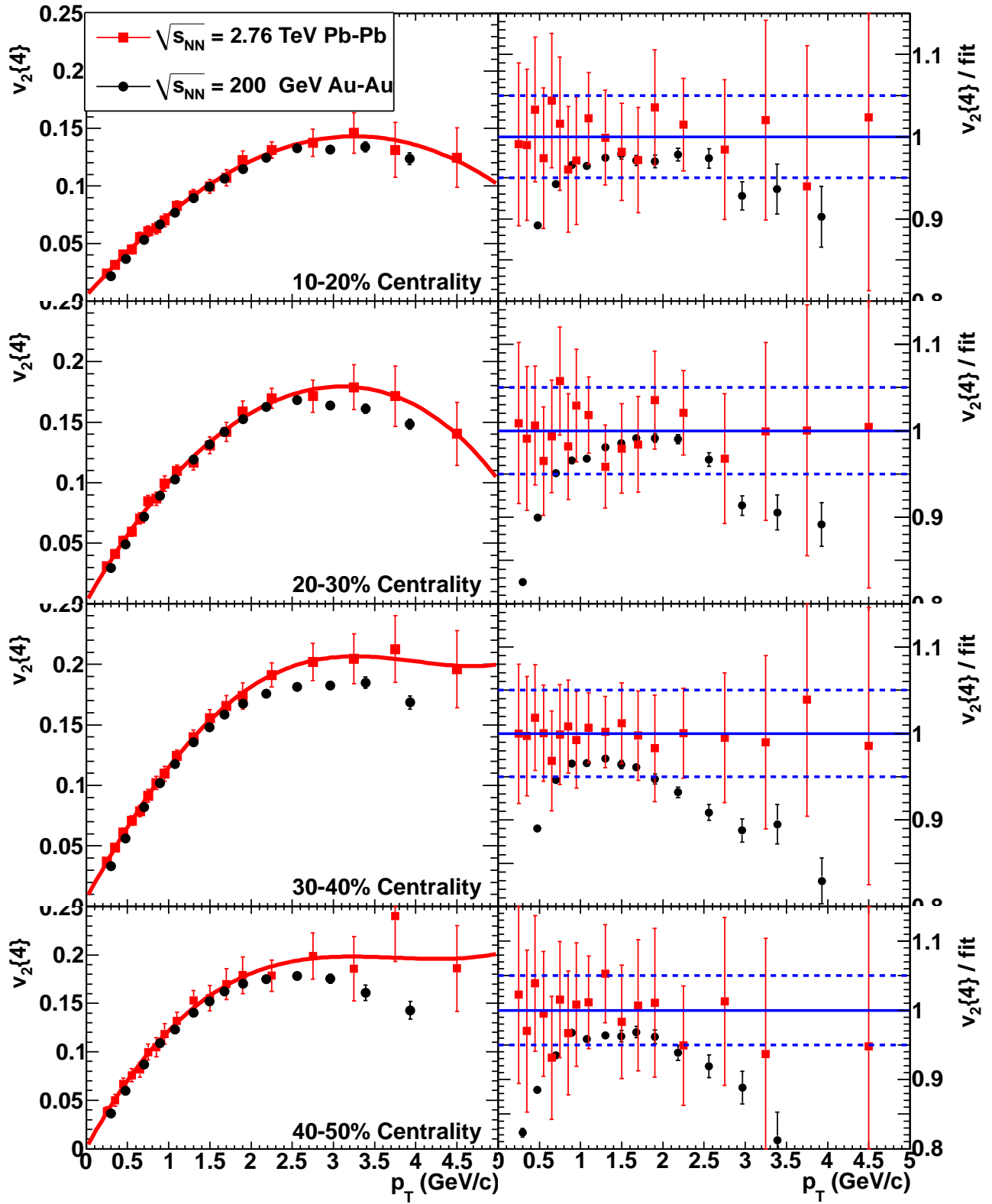}
\caption{Experimentally measured $v_{2}\{4\}$ as a function of transverse momentum $p_T$ for unidentified
  charged hadrons for centrality selections 10-20\%, 20-30\%, 30-40\%,
  and 40-50\% from top to bottom.  Published results from the ALICE experiment at
  2.76 TeV~\cite{Aamodt:2010pa} and the STAR experiment at 200 GeV~\cite{Abelev:2008ed} are shown.
  The right panels show the
  two data sets divided by a common fourth-order polynomial fit to the
  ALICE data points in each separate centrality selection.
\label{fig_data_compare2}}
\end{figure*}

The recently completed first LHC heavy ion run at  $\sqrt{s_{NN}} = 2.76$ TeV
has provided an enormous increase in collision energy over the top RHIC center-of-mass energy of $\sqrt{s_{NN}} = 200$ GeV. 
One of the more intriguing results is the
measurement from the ALICE collaboration of the elliptic flow $v_2$ as a function of transverse momentum $p_T$ for
inclusive unidentified charged hadrons ($h^{+} +
h^{-}$)~\cite{Aamodt:2010pa} as shown in
Figures~\ref{fig_data_compare} and ~\ref{fig_data_compare2}.  

In comparing their measured $v_{2}\{4\}$ (a
technique for measuring $v_{2}$ via four-particle cumulants~\cite{Borghini:2001vi}), with previous measurements at lower energies
at RHIC~\cite{Abelev:2008ed,Kumar:2011de}, a striking agreement is seen.
The optimal approach to quantify the level of agreement between the values of $v_2(p_T)$
measured at LHC and RHIC would employ a full statistical and systematic
uncertainty constraint fit following the methods developed in Ref.~\cite{Adare:2008cg}.
However, we note that in fitting a fourth-order polynomial to  the published ALICE data,  which
include {\em only} statistical uncertainties, we obtain a
$\chi^{2}/d.o.f. = 3.4 / 13$ corresponding to a p-value = 0.996; any inclusion of
systematic errors would only decrease the $\chi^2$ and increase the p-value.
The very low values of $\chi^2$ result from the 
fourth-order cumulant method which produces statistical
correlations between the data points~\cite{Raimond}.  Additionally, other sources of
systematic uncertainties are not fully quantified for either ALICE or
STAR data sets.  Regardless, as shown in the lower panel of
Figure~\ref{fig_data_compare} and the right panels of Figure~\ref{fig_data_compare2}, the agreement 
 between the 2.76 TeV and 200 GeV data is at the level of a few
percent from $p_{T} = 0.5$-2.5~GeV, with larger deviations possible at lower and higher $p_T$.
For the 20-30\% centrality bin shown in
Figure~\ref{fig_data_compare}, the agreement persists at the 5\% level down to energies at least as
low at 39 GeV.

The PHENIX experiment has
published data in Au-Au collisions at $\sqrt{s_{NN}}=$ 200 GeV on $v_{2}$ using the event plane method~\cite{Adare:2010ux} that
reveals a similar agreement~\cite{Lacey:2010ej}.  However, caution is
warranted regarding too fine a level of agreement or disagreement
on account of the different sensitivities of the methods to flow
fluctuations, which are
appreciable at the 10-20\% level in the final measured $v_{2}$~\cite{Ollitrault:2009ie}

\section{Perfect or Too Perfect}

In this paper, we discuss some of the implications and new questions
raised by this agreement of the $v_2$ measurements.
In doing so, we consider the following scenario.
Imagine that we could prepare a system with a specific spatial
distribution of energy density (in particular with a specific set of
spatial anisotropies)
and immediately evolve the system as a fluid with a specific fixed
value for $\eta/s$ to some appropriately large time.   Let the medium
have an Equation of State (EOS) where the speed of sound is
independent of temperature -- for example an ideal non-interacting gas with $\epsilon  = p/3$.  After the long
evolution time, the system immediately breaks up into hadrons via the Cooper-Frye freeze out mechanism~\cite{Cooper:1974mv}.  One might
expect that the momentum anisotropy (e.g. $v_2$) as a function of $p_T$
for hadrons would
be the same regardless of the initial energy density scale.  To be specific, if the central energy density were
a factor of four higher, but the spatial distribution of that larger energy density were identical, then despite larger pressure
gradients in all outward directions, the {\em relative} pressure gradients in different directions would be the same.
Thus, if one speculates that the fluid created at RHIC is ideal
($\eta/s = 0$) or nearly ideal ($\eta/s = 1/4\pi$) and the same is true at the LHC, perhaps this explains the near identical nature of the experimental data.
A number of assumptions (some more justified than others) went into the above scenario.  We examine some of
these assumptions within the context of various models and speculate
on what conclusions might be drawn.
Useful reviews of recent work on viscous hydrodynamics can be found in Refs.~\cite{Heinz:2009xj,Teaney:2009qa,Romatschke:2009im,Hirano:2008hy}.
%
%
%

\begin{figure}[ht]
\centering
\includegraphics[width=0.5\textwidth]{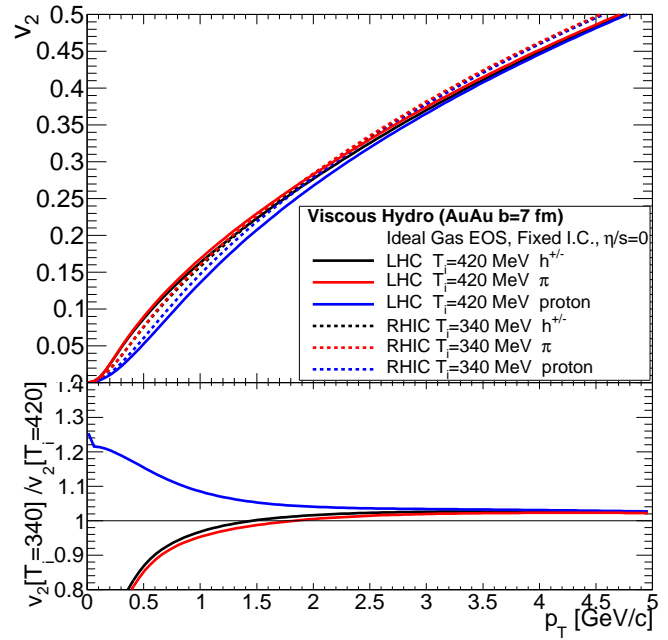}
\caption{Viscous hydrodynamic results using an ideal gas EOS and
$\eta/s = 0.001$.  Shown are the $v_{2}$ for unidentified hadrons, pions, and
  protons as a function of $p_T$.  The solid (dashed) lines are for the $T_{i}
  = 420 (340)$ MeV, and the lower panel shows the ratio of $v_2$
  values from the two cases.
\label{fig_rhiclhc_ideos}}
\end{figure}

First, we use the publicly available viscous hydrodynamic model of Romatschke and
Luzum~\cite{Luzum:2008cw, Romatschke:2007mq} to test this simple scenario.  
We fix the initial conditions for Au-Au collisions with impact parameter $b$ = 7 fm using
Monte Carlo Glauber results for binary collision positions~\cite{Luzum:2008cw}.  We use an
ideal gas EOS, $\eta/s = 0.001$ (not exactly zero for numerical
stability reasons and known to reproduce the results of ideal
hydrodynamical calculations), and an isothermal freeze-out temperature of 140~MeV
followed by Cooper-Frye hadronization and resonance decays.  We
model two cases, one with initial temperature $T_{i} = 340$~MeV and a
second with $T_{i} = 420$~MeV, as estimates for initial RHIC and LHC
temperatures respectively~\cite{Luzum:2009sb}.  We use identical
initial spatial distributions for both cases (with just a rescaling of
the energy density for the appropriate initial temperature), in order
to allow us to isolate the other dynamical effects.

\begin{figure*}[ht]
\centering
\includegraphics[width=0.42\textwidth]{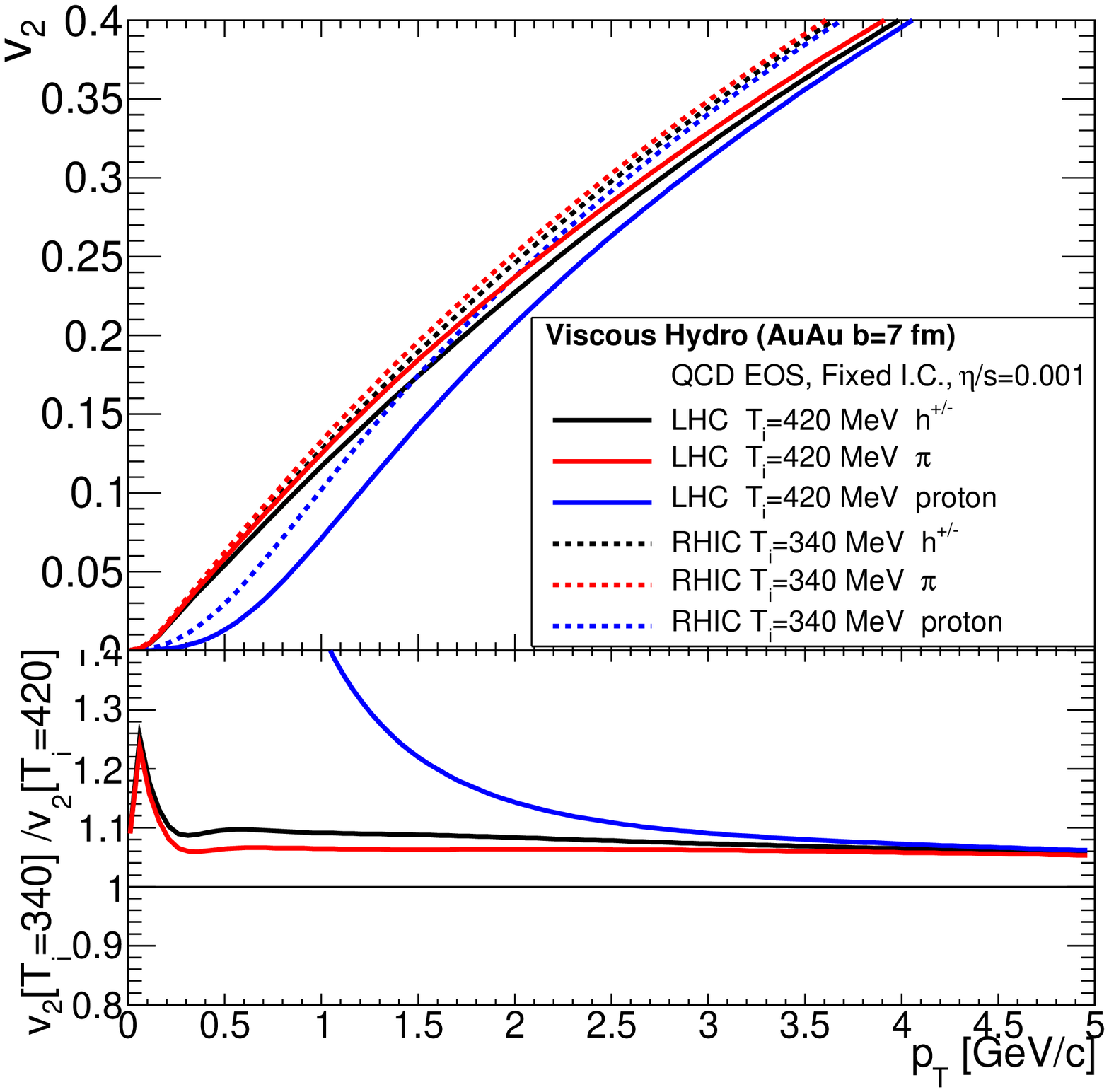}
\includegraphics[width=0.42\textwidth]{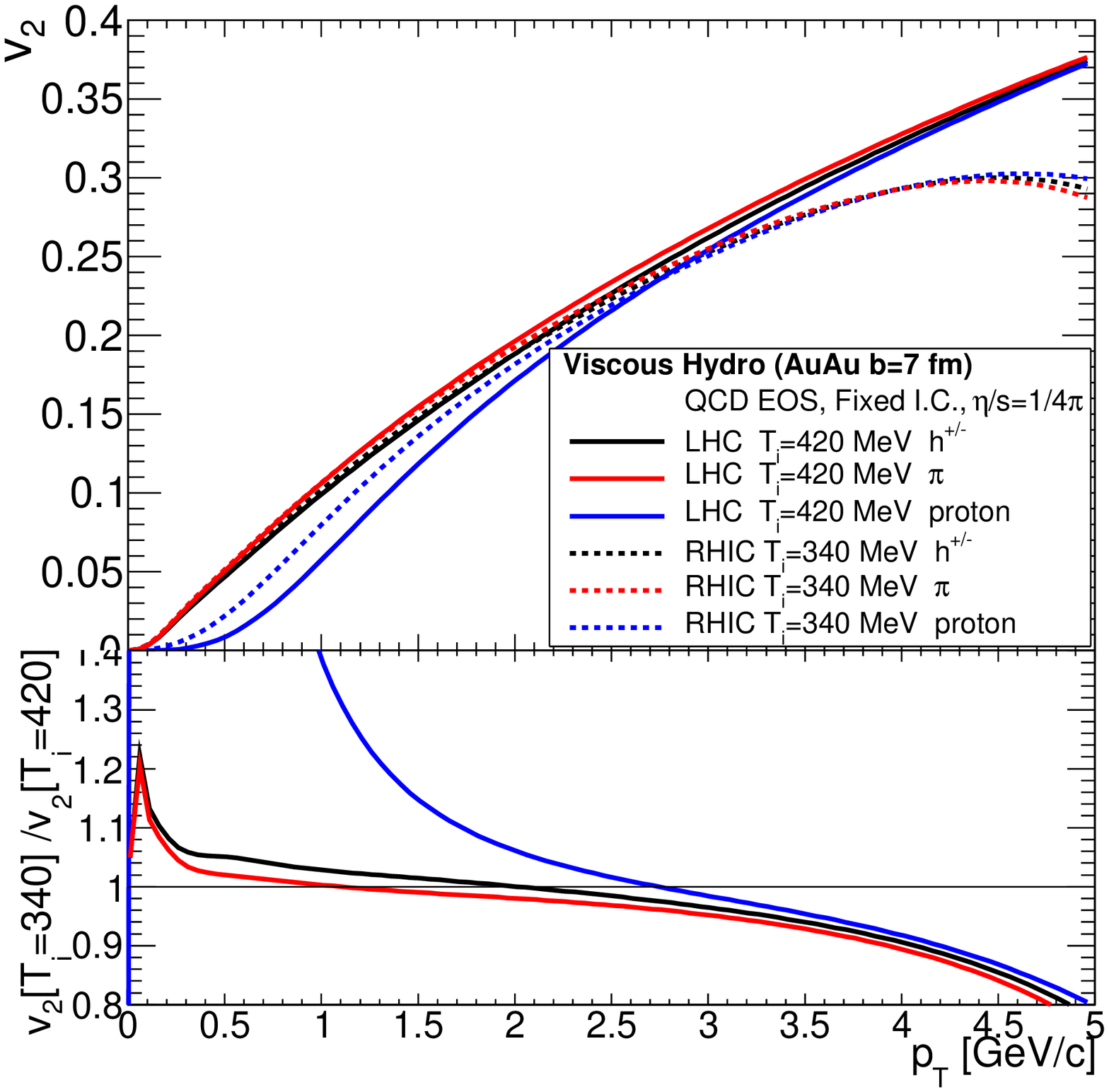}
\caption{Viscous hydrodynamic results using an EOS determined from lattice QCD and
$\eta/s = 0.001$ (left panel) and $\eta/s = 1/4\pi$ (right panel).  
Shown are the $v_{2}$ for unidentified hadrons, pions, and
  protons as a function of $p_T$.  The solid (dashed) lines are for the $T_{i}
  = 420 (340)$ MeV, and the lower panels show the ratio of $v_2$
  values from the two cases.
\label{fig_rhiclhc_qcdeos}}
\end{figure*}

The hydrodynamic results are shown in
Figure~\ref{fig_rhiclhc_ideos}.  As expected, the pions (and thus the
unidentified hadrons which are dominantly pions) have the same $v_2$ within a few percent for
$p_{T} > 1$~GeV.  However, the proton $v_{2}(p_{T})$ pattern appears shifted
out in $p_T$ for the higher initial temperature case. This is most
likely due to the larger radial boost mapping onto the momentum
distribution of protons.  One speculation is that the deviation at low $p_T$ for pions may be
from resonance decay contributions.
However, we have checked the primary (without decay contribution)
$v_2$ for pions and similar differences between the two
initial temperature cases is observed.
%
%
%
%

%
%
%
%

Second, we extend this question to the case of the QCD EOS (using the
lattice QCD inspired EOS from Ref.~\cite{Laine:2006cp})
%
%
%
and for $\eta/s = 0.001$.  The left panel in Figure~\ref{fig_rhiclhc_qcdeos} shows the results from
this calculation again with two initial temperatures of 420 and 340 MeV.
Despite the variation  in the speed of sound $c_s$ as a
function of temperature due to the EOS, we find a similar qualitative result.  For
the range $p_{T} = 0.5 - 3.5$~GeV the pions, and therefore the
unidentified charged hadrons, have a $v_2$ that is the same within
approximately 5\%.  It is notable that this $v_2$ signal is actually
somewhat stronger for the lower $T_{i} = 340$ MeV.  
However, the proton $v_{2}(p_{T})$ pattern again appears shifted
out in $p_T$ for the higher initial temperature case. 

The right panel of
Figure~\ref{fig_rhiclhc_qcdeos}
shows the results for the same QCD EOS but now with $\eta/s = 1/4\pi$.
In this case there is a noticeable difference at the highest $p_T$
for all particles with the lower temperature case having a lower $v_2$. 
%
%
%
%
%
%
This observation suggests that even a minimal viscosity plays a significant
role in limiting the growth of $v_2$ with $p_T$ for  $p_T > 3$~GeV at RHIC energies,
while the significantly higher energy densities at the LHC extend the regime
where inertial forces dominate dissipation.  We have run the $T_{i} =
420$ MeV case to higher hadron $p_{T}$ and find that the $v_{2}$ values do
start to saturate for $p_{T} > 6.5$ GeV  at a level
near $v_{2} \approx 0.44$.
See Section~\ref{Sec:Sensitivity} for a discussion of
other effects that must also be considered in this momentum regime.
It is possible a hint of this is seen in the experimental 200~GeV data in
Figure~\ref{fig_data_compare2} in all centrality selections; but current uncertainties preclude any
strong conclusion.  
Again, the proton $v_2$ pattern is shifted to higher
$p_T$, leading to a rising ratio for the lower to the higher
temperature case at $p_{T} < 2.0$~GeV.
Thus, the measured $v_2$ for
protons at the LHC is much anticipated.

Note that we have used smooth Glauber initial conditions for fixed
impact parameter b = 7 fm and no hadron
re-scattering after freeze-out of the fluid in these calculations.
Therefore, these quantitative calculations should not be compared
directly with the experimental data, but rather reveal the qualitative trends
with changing initial temperature.  

\section{Sensitivity to $\eta/s$}
\label{Sec:Sensitivity}

%
%
%
%
%

A crucial experimental question is the sensitivity of $v_2(p_T)$ to the value of
$\eta/s$. A corresponding critical question for theory is the 
assumption employed by nearly all hydrodynamic calculations to date
that $\eta/s$ is independent of temperature. In this section we 
examine these issues.

\begin{figure}[t]
\centering
\includegraphics[width=0.5\textwidth]{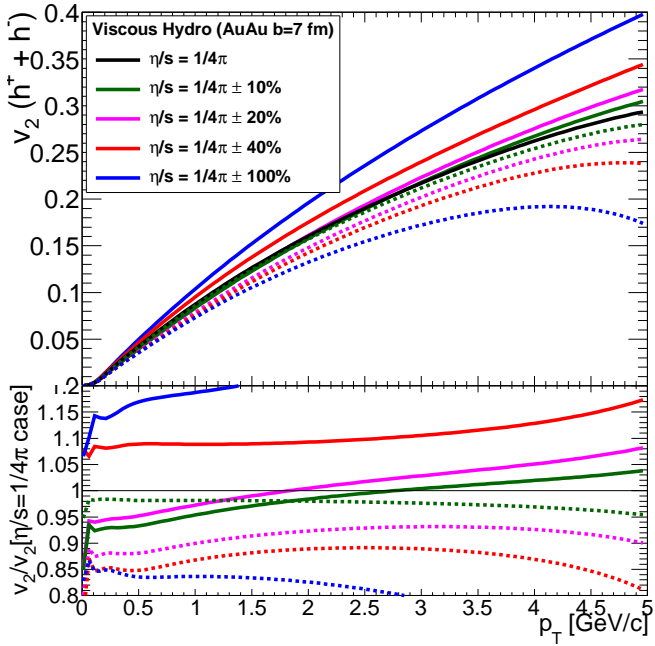}
\caption{Viscous hydrodynamic results using QCD EOS for Au-Au $b$=7 fm
  events, initial temperature T = 340 MeV, and with varying $\eta/s$.  Shown are the $v_{2}$ for
  unidentified charged hadrons as a function of $p_T$.  The black curve
  is for $\eta/s = 1/4\pi$.  The other colors correspond to changes of
  10, 20, 40, 100\% in $\eta/s$.  The solid (dashed) lines correspond
  to a decrease (increase) in the ratio.  The lower panel shows the
  rato of all curves to the $1/4\pi$ case.
\label{fig_vary_etaos}}
\end{figure}

We have calculated the $v_2(p_T)$  for
unidentified charged hadrons for RHIC temperatures with different
variations in $\eta/s$ as shown in Figure~\ref{fig_vary_etaos}.  The
black solid curve corresponds to $\eta/s = 1/4\pi$, the blue solid curve to
$\eta/s \approx 0$, and the blue dashed curve to $\eta/s = 2/4\pi$.
Note that we have not re-normalized the initial entropy density of the
medium to maintain a fixed final particle multiplicity or mean
transverse momentum as done in other works -- for example see Ref.~\cite{Shen:2011kn}.
For 10\% variations in $\eta/s$ (green curves), there is very little
change in the $v_2$ values (less than 5\%). In fact,
for these small 10-20\% increases in $\eta/s$,  $v_2$ shows a slight
increase for $p_T$ below $\sim 2$~GeV. 

It is clear that even a $\pm 40$\% change in $\eta/s$ (red curves) results in only
a 10\% variation in $v_2$.  Furthermore, it is important to
note that this 10\% variation is nearly independent of $p_T$ from 0-5~GeV.  
Many comparisons of experimental data and viscous
hydrodynamics focus on the differences at higher $p_T$ because the
theoretical calculations with different values of $\eta/s$ {\em visually} split (as shown on a linear scale).  
However, this region in $p_T$ is also
sensitive to other effects such as the implementation of the departure $\delta f(p)$
from the equilibrium distribution $f_0(p)$~\cite{Dusling:2009df}
and path length dependent jet energy loss (see for
example~\cite{Jia:2011pi}).

Given the relative insensitivity of $v_2(p_T)$ to the precise value
of $\eta/s$, it is important to identify analysis procedures that maximize 
the sensitivity to this important quantity.
Plotting the ratio of  $v_2(p_T)$  between experiment measurements
and theoretical calculations reveals important discriminating power at low $p_T$.  It is at
this low $p_T$ that the experimental statistical uncertainties are
smallest, and the systematic uncertainties 
(which require further attention to fully quantify) 
are percentage uncertainties and thus largely
independent of $p_T$..

\begin{figure}[t]
\centering
\includegraphics[width=0.5\textwidth]{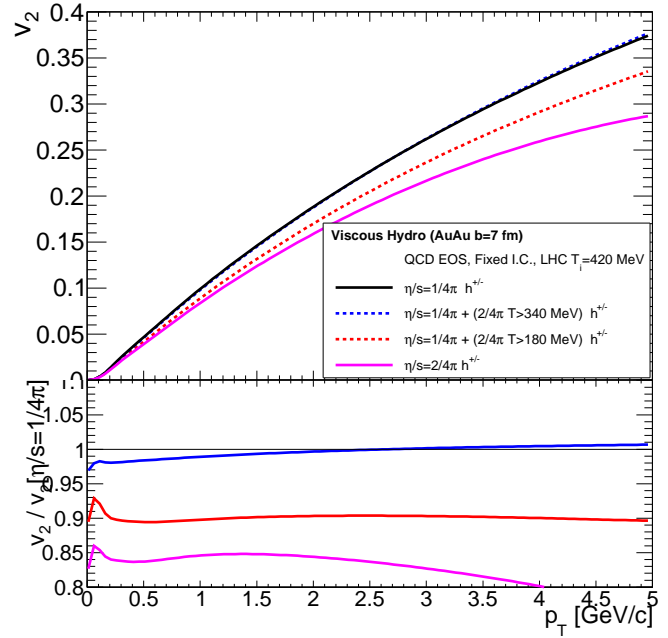}
\caption{Viscous hydrodynamic results for $v_2$ versus $p_T$ for
  unidentified charged hadrons using a QCD EOS and initial
  temperature T = 420 MeV.  We compare results with fixed $\eta/s =
  1/4\pi$, $\eta/s = 2/4\pi$, and two cases with a temperature
  dependent value where $\eta/s = 1/4\pi$ for $T<340$~MeV and $\eta/s
  = 2/4\pi$ for $T > 340$ MeV or $\eta/s = 2/4\pi$ for $T>180$ MeV.
The lower panel shows the
  rato of all curves to the constant $\eta/s = 1/4\pi$ case.
\label{fig_varetaos_step}}
\end{figure}

In most hydrodynamic calculations the specific $\eta/s$ value which is used is often referred to as an
average.  However, it is not simple to define what type of
average this quantity is, as it is weighted in some complicated fashion over the space and time evolution.  
More realistically, it should be regarded as 
simply a fixed value used to obtain  a first-order estimate.  In this
simple picture, the RHIC and LHC mediums would require values for
$\eta/s$ that differ significantly less than 40\%.

We can examine the sensitivity to possible $\eta/s$ variations with temperature
by a straightforward procedure.  As a baseline, we have run with the
QCD EOS and an initial temperature of 420~MeV and constant values for
$\eta/s = 1/4\pi$ and also $\eta/s = 2/4\pi$.
For comparison we have implemented a simple step-function temperature dependence
where for $T < 340$~MeV $\eta/s = 1/4\pi$ and for $T > 340$ MeV $\eta/s =
2 \times 1/4\pi$ (i.e. 100\% larger).  We also consider a second case
where the step-function occurs near the transition temperature $T_{c} = 180$~MeV,
that is, $\eta/s = 1/4\pi$ for $T<T_{c}$ and $\eta/s = 2 \times 1/4\pi$ for $T>T_{c}$.
The results are shown in Figure~\ref{fig_varetaos_step}.

For the first case, the results show essentially no difference in $v_2(p_T)$ over
the $p_T$ range shown.  The T=340-420 MeV range is that
explored in the early times for the LHC created medium, and then after
cooling it explores the same temperature range as that for the RHIC
created medium. 
Notably, there is no measurable difference in $v_2(p_T)$  for a factor of two change
in the $\eta/s$ ratio for the early time higher temperature medium. 
Even if the step-function in the value of $\eta/s$ occurs at the much
lower temperature of $T_{c}=180$~MeV,  then only a 10\% decrease in $v_2$ at
all $p_T$ is seen.  The modest values of these changes is striking, 
and pose serious challenges to more precise extraction of transport
coefficients via such measurements. 

Recently a study exploring a
family of four temperature-dependent $\eta/s$ cases has been performed~\cite{Niemi:2011ix}, with a
change in $\eta/s$ at the transition temperature $T_{c} = 180$~MeV that
includes a possible 
sharp rise in $\eta/s$ just above the transition (for example in the case labeled HQ the
$\eta/s$ value rises to $10 \times 1/4\pi$
by $T > 400$ MeV).   While these very large values for $\eta/s$ call into question the
validity of the calculation in that parameter space~\cite{Huovinen:2008te}, 
even in this extreme comparison, only a modest (less than 15\%) change in the predicted
$v_2$ is found at LHC energies (specifically the so-labeled
LH-LQ to LW-HQ cases).  Qualitatively this is consistent with our finding where
our very small (by comparison) factor of two change located at a higher
temperature ($T>340$~MeV) has
very little impact.  Thus, it appears {\em premature} to decide whether the
higher temperature region explored at the LHC (before cooling and
evolving over the same temperature range as RHIC) has the same or
different $\eta/s$ (as predicted in some models -- for example~\cite{Hidaka:2008du}).
It is clear that more detailed studies of the temperature dependence
are necessary; work in that direction is underway~\cite{Shen:2011kn,Niemi:2011ix}.
A parallel critical area of research is recent work towards
calculating these dynamical properties (for example $\eta/s$ as a
function of temperature) for QCD on the lattice~\cite{Meyer:2010tt}.

\section{Initial Conditions}
\label{Sec:Eccentricity}

In all of the above discussion we have deliberately assumed that the initial spatial eccentricity
and thermalization times do not vary between RHIC and LHC energies
 in order to separately study the effects due to $\eta/s$ variations, the EOS and particle mass.
Since ideal hydrodynamics predicts that $v_2$ should be
proportional to the initial eccentricity, it is also important to investigate 
whether the initial spatial distribution at the LHC is the same as at
RHIC.   Another expected difference between these two energy regimes
is the equilibration time $\tau$. It is often stated
that the hydrodynamic matching to experimental data at RHIC indicate
rapid equilibration $\tau < 1.0-2.0$ fm/c.  
If in fact  the final $v_2$ pattern is sensitive to
the equilibration time, and $\tau$ is significantly smaller at the LHC than at RHIC,
the near-identity of the $v_2(p_T)$ data from RHIC and the LHC is deeply mysterious.
Of course, one solution to this mystery is that the input assumptions are wrong, 
as discussed in Ref.~\cite{Luzum:2008cw}, where it is argued that the purported  sensitivity to rapid
equilibration is incorrect
%
%

\begin{figure*}[th]
\centering
\includegraphics[width=0.45\textwidth]{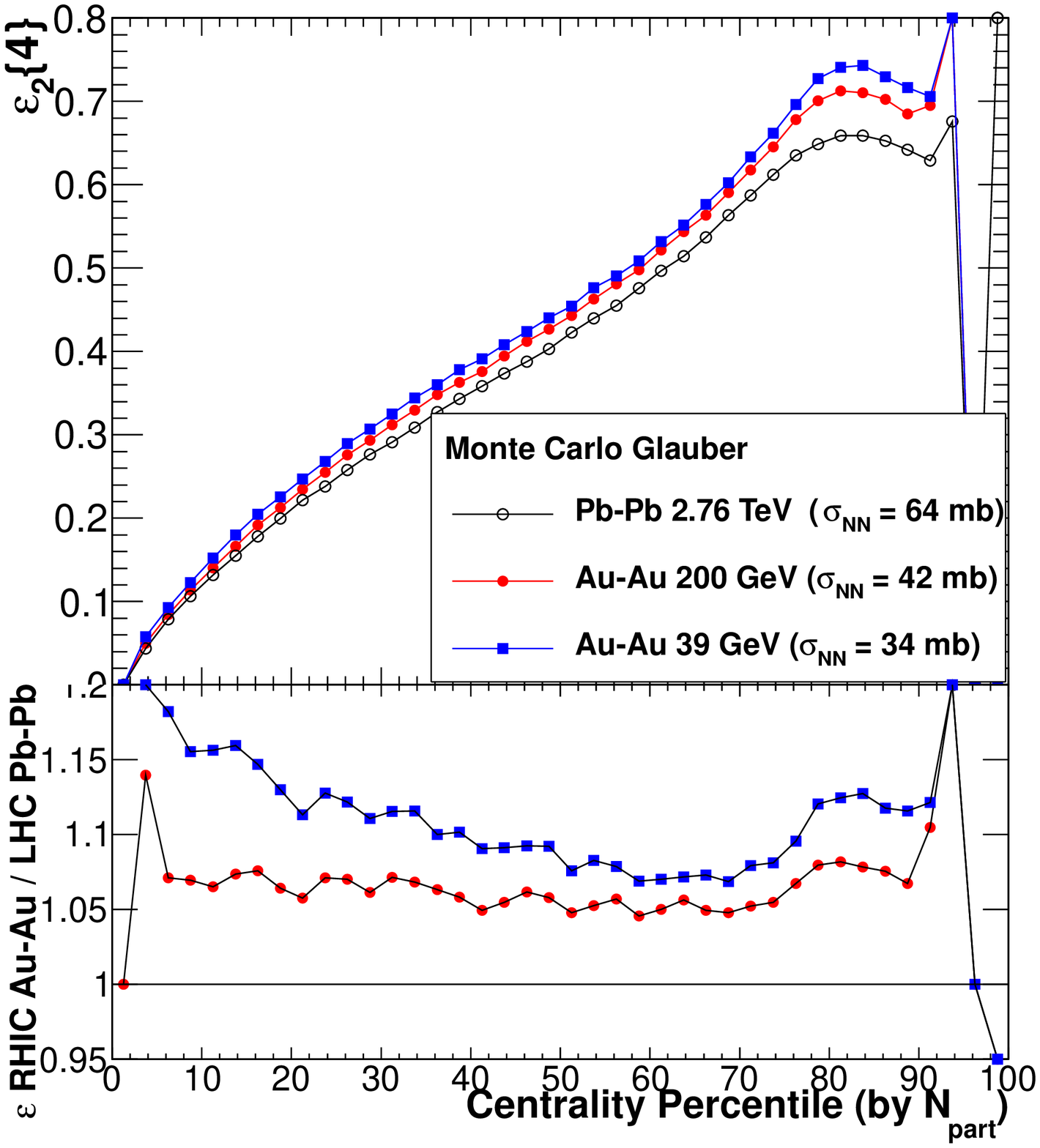}
\includegraphics[width=0.45\textwidth]{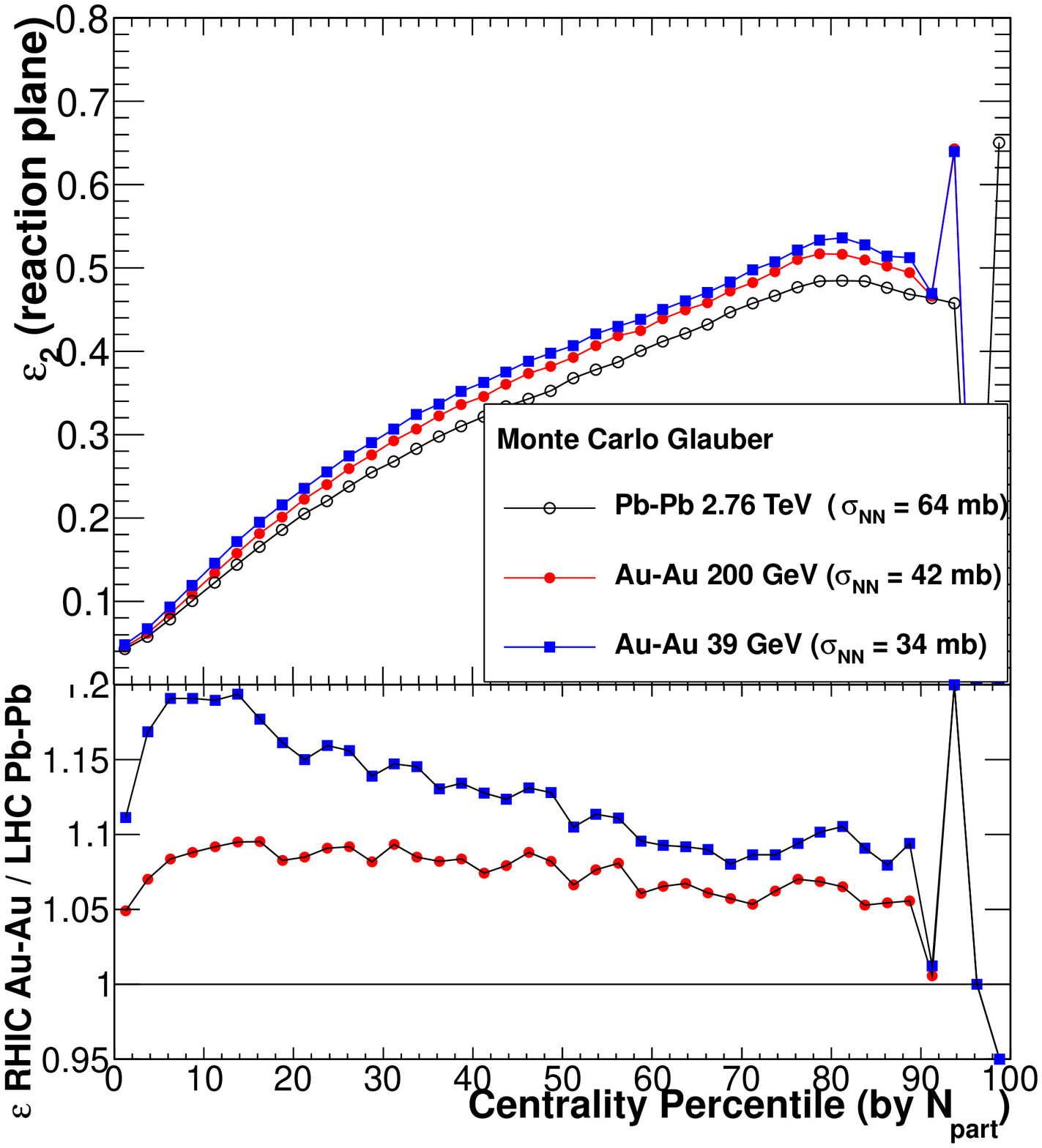}
\caption{Monte Carlo Glauber results for Au-Au collisions at 39 and 200 GeV
  and Pb-Pb collisions at 2.76 TeV.  The left panel shows 
  $\epsilon_{2}\{4\}$ with respect to the
reaction plane, 
and the right panel shows the
  RMS $\epsilon_{2}$ with respect to the participant plane, all as a
  function of collision centrality percentile (determined from the
  number of participating nucleons).  The lower panels show the
  ratio of the RHIC to LHC eccentricity values.  
\label{fig_glauber}}
\end{figure*}

In order to understand the potential range of variation of initial eccentricity,
we have examined  two of the currently used phenomenological models for
calculating the initial spatial eccentricity.
First, we use a
Monte Carlo Glauber framework~\cite{Alver:2008aq} to calculate the
different eccentricities as a function of collision centrality
percentile as determined by the number of participating nucleons ($N_{part}$).
The most appropriate basis of comparison to compare data sets are the 
centrality percentiles (e.g. 20-30\% of the total inelastic cross
section) being used by experiments; 
the eccentricity ratios show considerably more variation if comparisons are made
at fixed participant number or impact parameter.
%
%

As already mentioned in Section~\ref{sec:introduction}, there are multiple techniques for measuring $v_2$ that
give results varying by of order 20\%.  These variations are due to
the different influences of flow fluctuations and non-flow effects~\cite{Ollitrault:2009ie}.
If we assume that initial eccentricities $\epsilon_{n}$ translate into
momentum anisotropies $v_{n}$ in individual events, then
the most applicable for the $v_{2}\{4\}$ measurements is 
$\epsilon_{2}\{4\} = 
\left[ 
2 \left< \epsilon^{2} \right>^2
-
 \left<\epsilon^{4} \right> 
\right]^{1/4}$, 
which is shown in the left
panel of Figure~\ref{fig_glauber}.  
%
%
%
%
%
One observes an approximately 5\% (10\%) larger eccentricity for Au-Au at 200 (39)~GeV collisions
compared with Pb-Pb collisions at 2.76~TeV.  This 5\% difference
between 200~GeV and 2.76~TeV values was previously noted
in~\cite{Aamodt:2010pa}.

There are three important contributors to
these differences.  First, the simple difference of atomic mass between Au (197)  and Pb
(208).  Second, the nucleon-nucleon inelastic cross sections are significantly increasing
across the energy interval spanned by RHIC and the LHC, ranging from 
$34 \pm 3$ mb at RHIC 39 GeV, $42 \pm 2$ mb at RHIC 200 GeV, and $64 \pm 5$ mb
at LHC 2.76~TeV (although
this last value is not yet experimentally finalized).  
A larger cross section reduces the fluctuations in the number of participants, particularly
in the periphery of the interaction zone, which in turn reduces the magnitude of 
$\epsilon_{2}\{4\}$, 
as seen in the left panel of  Figure~\ref{fig_glauber}.
%
%
Third, and related
to the second, is that the larger  total A-A inelastic cross section 
at the LHC changes the mapping of centrality percentile to
impact parameter range.

In the right panel we show the $\epsilon_{2}$ values with
respect to the reaction plane.  This shows a slight larger 8\% (15\%)
differences between 200 (39) GeV and 2.76 TeV.  When comparing the corresponding values of
$\epsilon_{2}$ with respect to the reaction plane and $\epsilon_{2}\{4\}$, there is an increasing
level of disagreement for more peripheral reactions, which is understood to result from 
non-Gaussian fluctuations that reflect the underlying Poisson fluctuations 
from discrete nucleons~\cite{Voloshin:2007pc,Alver:2008zza,Alver:2008aq}. 
More central collisions quickly approach the limit of Gaussian fluctuations
where the two estimates agree.
%
%
%
However, for some hydrodynamic model calculations the role of (presumably real) 
fluctuations has been greatly reduced in spite of using Monte Carlo
Glauber initial conditions by keeping the events fixed with respect to
the reaction plane and {\em averaging} over many events to create smooth
initial conditions as input.  
%
%
Note that this is exactly what is done for the hydrodynamic model
comparisons in~\cite{Hirano:2010je}, where even for
40-50\% central events this results is a 10\% underestimate in the
average eccentricity.  The effect is even larger (more than a 25\%
underestimate) for the 70-80\% centrality. Caution is therefore
warranted in making such comparisons with hydrodynamic models using
this initialization and data.  A useful discussion of this point is
made in Ref.~\cite{Song:2011hk}, and may point to the need for running
multiple hydrodynamic events on individually fluctuating initial conditions~\cite{Schenke:2010rr}.
%
%

We have also made similar comparisons for the Color Glass Condensate (CGC)
model motivated initial condition calculation.  This class of Monte
Carlo calculations start with the same Glauber model described above to determine
the participants in a collision.
However, a spatial region with say 20 participating nucleons is no longer
assumed to produce ten times the energy density as a region with 2
participating nucleons.  This is due to the assumed saturation in the number
of low-$x$ gluons, whose wavefunctions overlap across the longitudinal
extent of the nucleus.  These calculations produce eccentricities that
are larger than those from the pure Monte Carlo Glauber based on
participating nucleons alone~\cite{Drescher:2006pi}.

\begin{figure*}[th]
\centering
\includegraphics[width=0.45\textwidth]{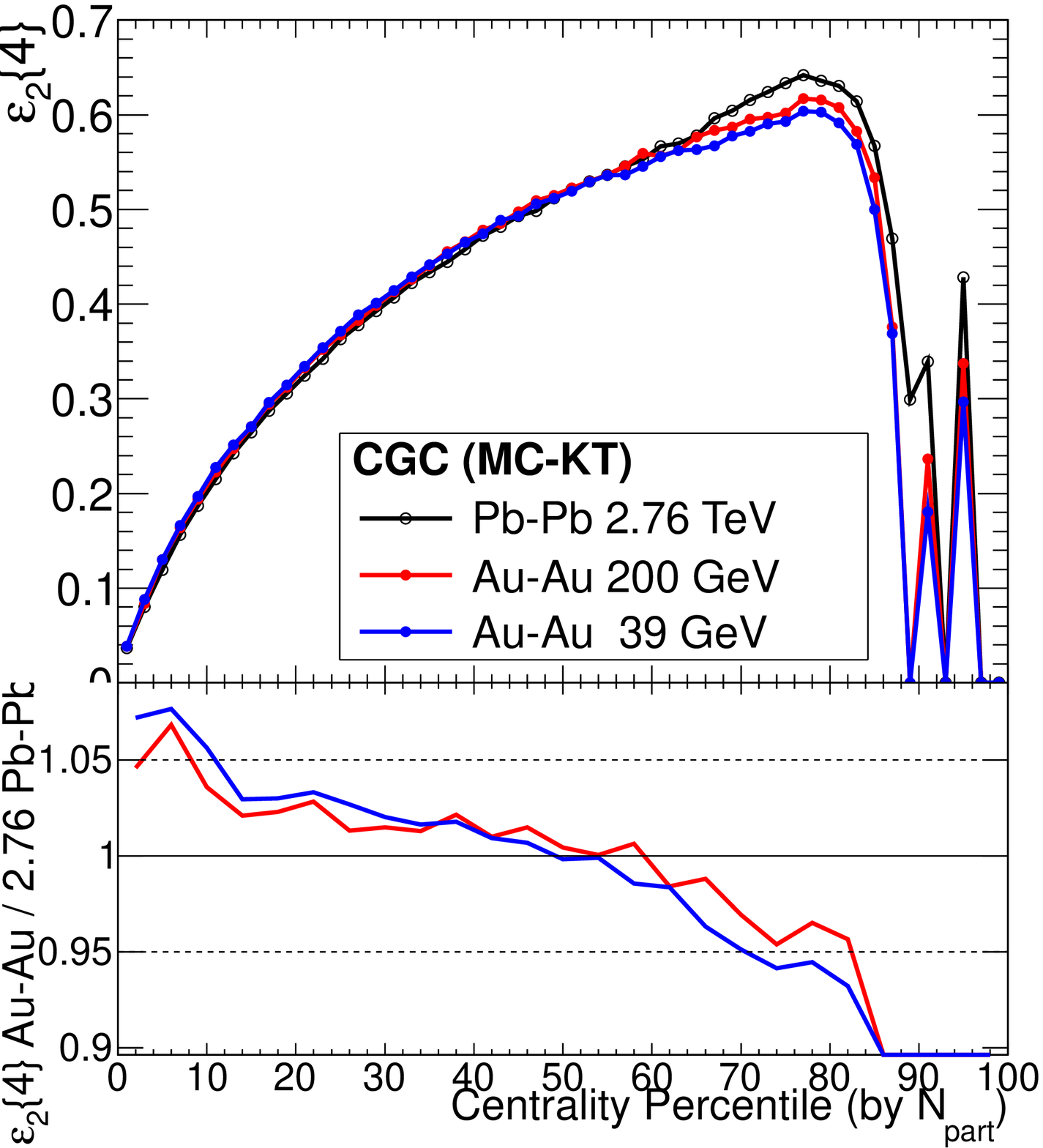}
\includegraphics[width=0.45\textwidth]{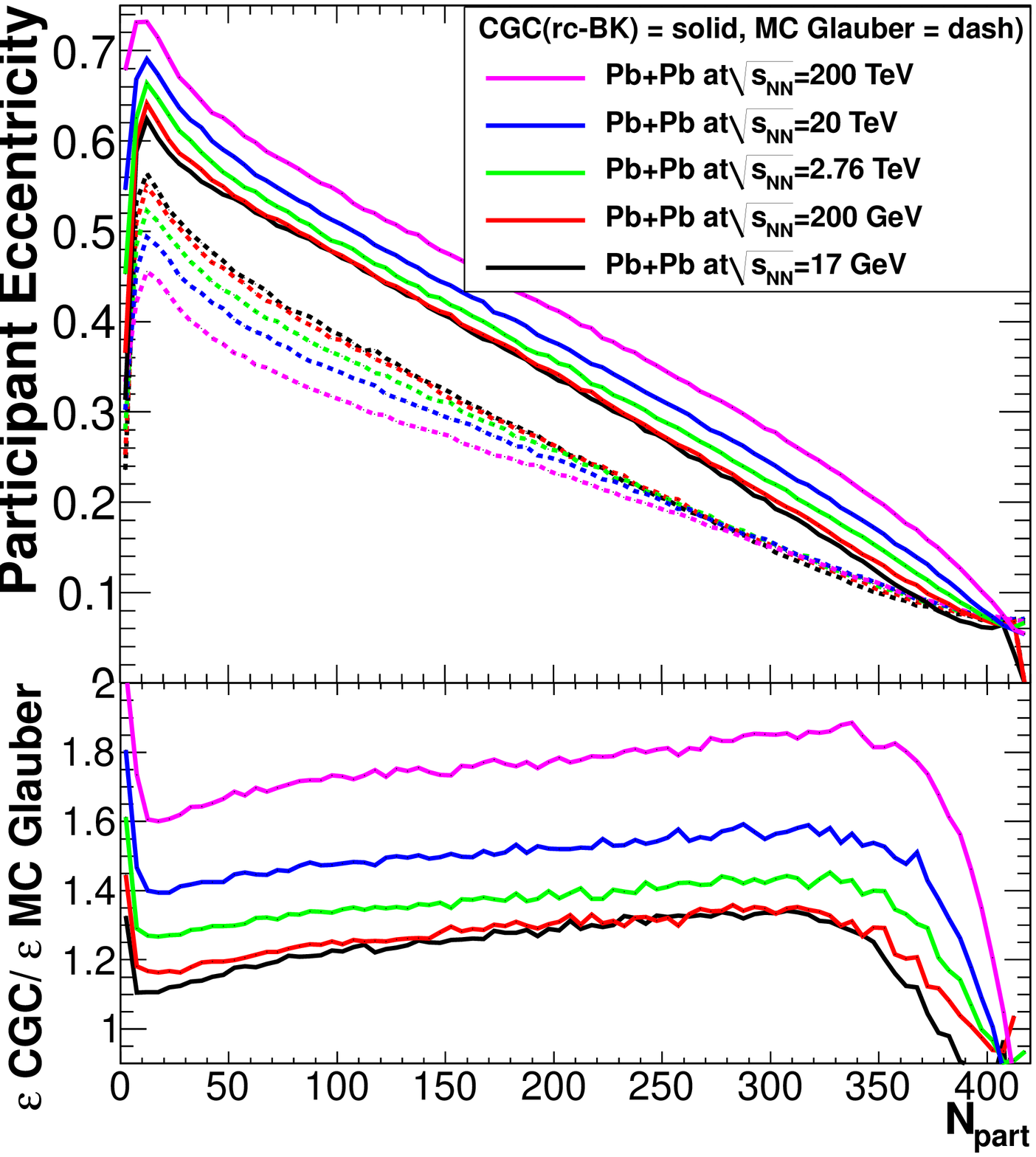}
\caption{The left panel shows the rc-BK CGC calculation for $\epsilon_{2}\{4\}$
  with respect to the participant plane for Au-Au at 39
  and 200  GeV and Pb-Pb at 2.76 TeV as a function of collision centrality
  percentile.  The lower left panel shows the ratio of LHC to RHIC
  eccentricities.
The right panel shows the CGC and Monte Carlo Glauber mean
eccentricities for Pb-Pb collisions over a broad range of collision
energies. The ration of CGC eccentricities to those for the
Monte Carlo Glauber are shown in the lower panel.
\label{fig_cgc}}
\end{figure*}

Here we utilize the rc-BK option~\cite{Albacete:2010ad} for the CGC
initial conditions.  Figure~\ref{fig_cgc} shows $\epsilon_{2}\{4\}$
with respect to the participant plane as a function of
centrality percentile for Au-Au at 39 and 200~GeV and Pb-Pb at 2.76~TeV.  One
observes in the lower panel that the eccentricities agree within less
than 2\% for much of the centrality range (10-70\%).  At first this result
seems counterintuitive since for higher collision energies, one
should be probing lower-$x$ gluons, saturation effects should increase,
and the eccentricity should increase relative to the Monte Carlo
Glauber case.
%
%
%
One does see a slight hint of this in that the Glauber
difference between 200 GeV and 2.76 TeV shown in the left panel of Figure~\ref{fig_glauber} is 5\%
and in this case it is reduced to 2\%.

In order to test this picture further, we have calculated the mean
eccentricity with respect to the participant plane for Pb-Pb
collisions for a large range in collision energies (17 GeV - 200 TeV).
Shown in the right panel of Figure~\ref{fig_cgc} are the resulting
eccentricities for the rc-BK  CGC case (solid lines) and the pure
Monte Carlo Glauber (dashed lines) as a function of the number of
participating nucleons.  In the lower right panel we calculate for each
collision energy the ratio of CGC to Glauber eccentricity.   As
expected, the larger the collision energy, the larger the modification
in the number of gluons and thus the larger the increase in
eccentricity for the CGC case.  However, the relative increase in
eccentricity is only $\approx$ 10\% from 200 GeV to 2.76 TeV. 
Curiously, this increase in eccentricity from saturation effects is largely canceled by the
decrease in the underlying Monte Carlo Glauber eccentricity and in the
mapping of centrality percentiles for Au-Au and Pb-Pb,  as shown in the
left panel.
%
%
%
%

One striking feature is that between  17 and 200~ GeV the CGC predicted $\sim20$\%
increase of eccentricity relative to the Monte Carlo Glauber changes only very slightly.
 In fact, the authors of the model note
that for energies below 130 GeV the formalism breaks down since one is
probing gluons with $x > 0.01$.  This raises the interesting
possibility that below some energy, there should be a transition in
initial conditions from the larger CGC calculated eccentricities to
the lower Monte Carlo Glauber values.  Data from the CERN-SPS heavy
ion fixed target program have not led to a definitive answer
regarding $v_2$ at these energies due to different methods,
centralities, and varying baryon contributions to unidentified
hadrons~\cite{Alt:2003ab}.  With additional data from the RHIC energy
scan, it will be very interesting to see how and if the $v_2$
decreases as the center-of-mass is lowered  (as predicted by many due to viscous effects in
the hadronic re-scattering stage).  The possible difference in initial
eccentricity will require careful study (since CGC effects are not
expected to play a role at these energies), and may be masked by 
other effects with a larger dynamics range, such as the 
(also poorly known) modifications to the 
EOS at finite baryon chemical potential.

\section{Averages and Knudsen Number Method}

The above discussion regarding the sensitivity to various parameters
and their temperature dependence raises the question of whether
simple models with a single $\eta/s$ value, constant speed of sound, and
a single characteristic temperature can 
go beyond simple dimensionful scaling arguments to both
capture the key physics and provide a basis for quantitative
extraction of key dynamical parameters such as $\eta/s$. 
For example, it has been proposed that one can extract $\eta/s$
from a fit to  $v_2$ (integrated over all $p_T$) versus particle density
using a simple scaling {\em ansatz} in the Knudsen number $K \equiv \lambda/{\bar R}$
~\cite{Drescher:2007cd} (here $\lambda$ is the mean free path for
momentum transport and $\bar R$ is a characteristic system size discussed below). 
Previously, we
had investigated some of the underlying assumptions for this model
and found important ambiguities in the parameterizations and
sensitivities~\cite{Nagle:2009ip}.  Despite these findings, the
quantitative values from~\cite{Drescher:2007cd} are frequently quoted
as reliable estimates for $\eta/s$.

\begin{figure*}[th]
\centering
\includegraphics[width=0.45\textwidth]{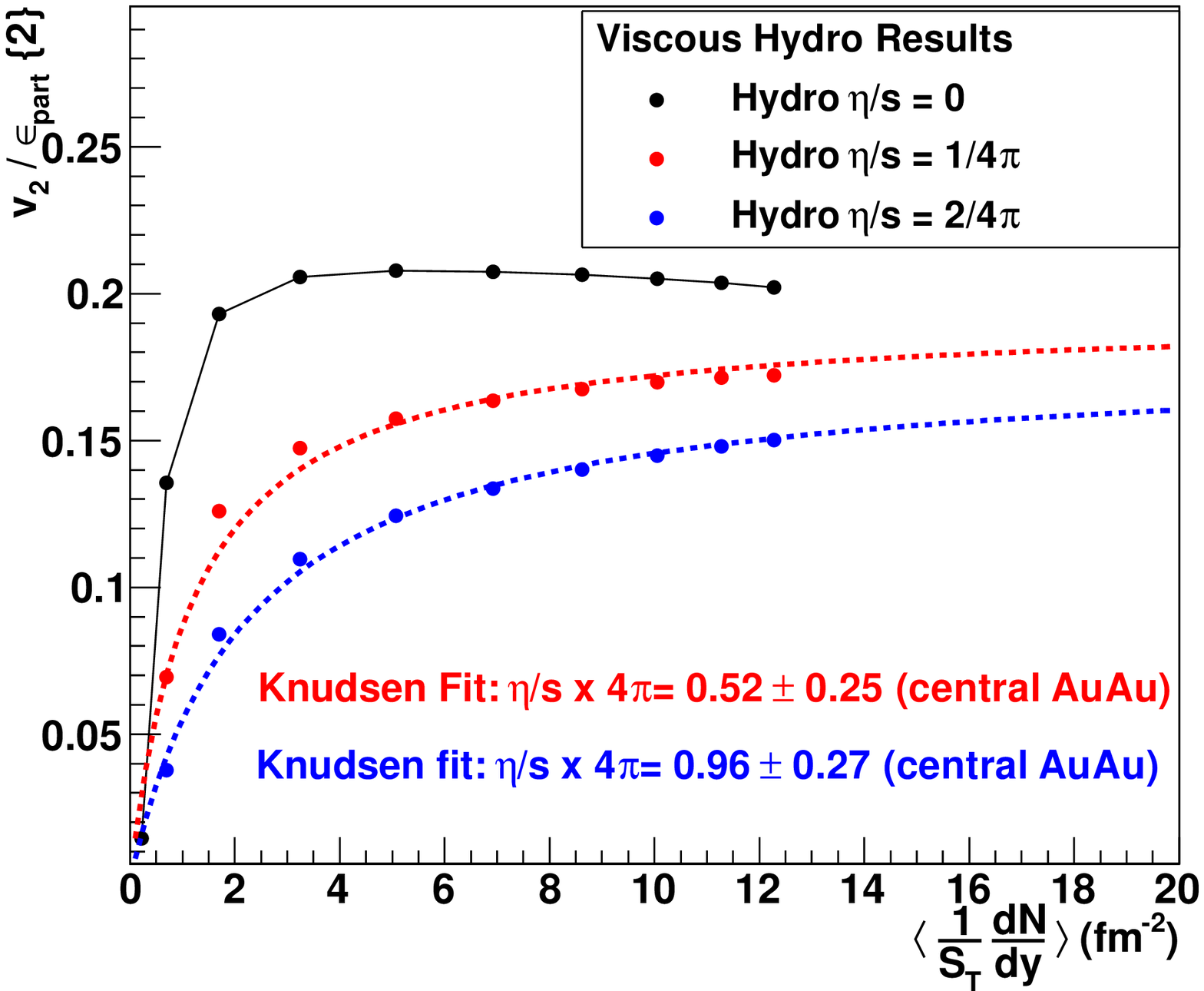}
\includegraphics[width=0.45\textwidth]{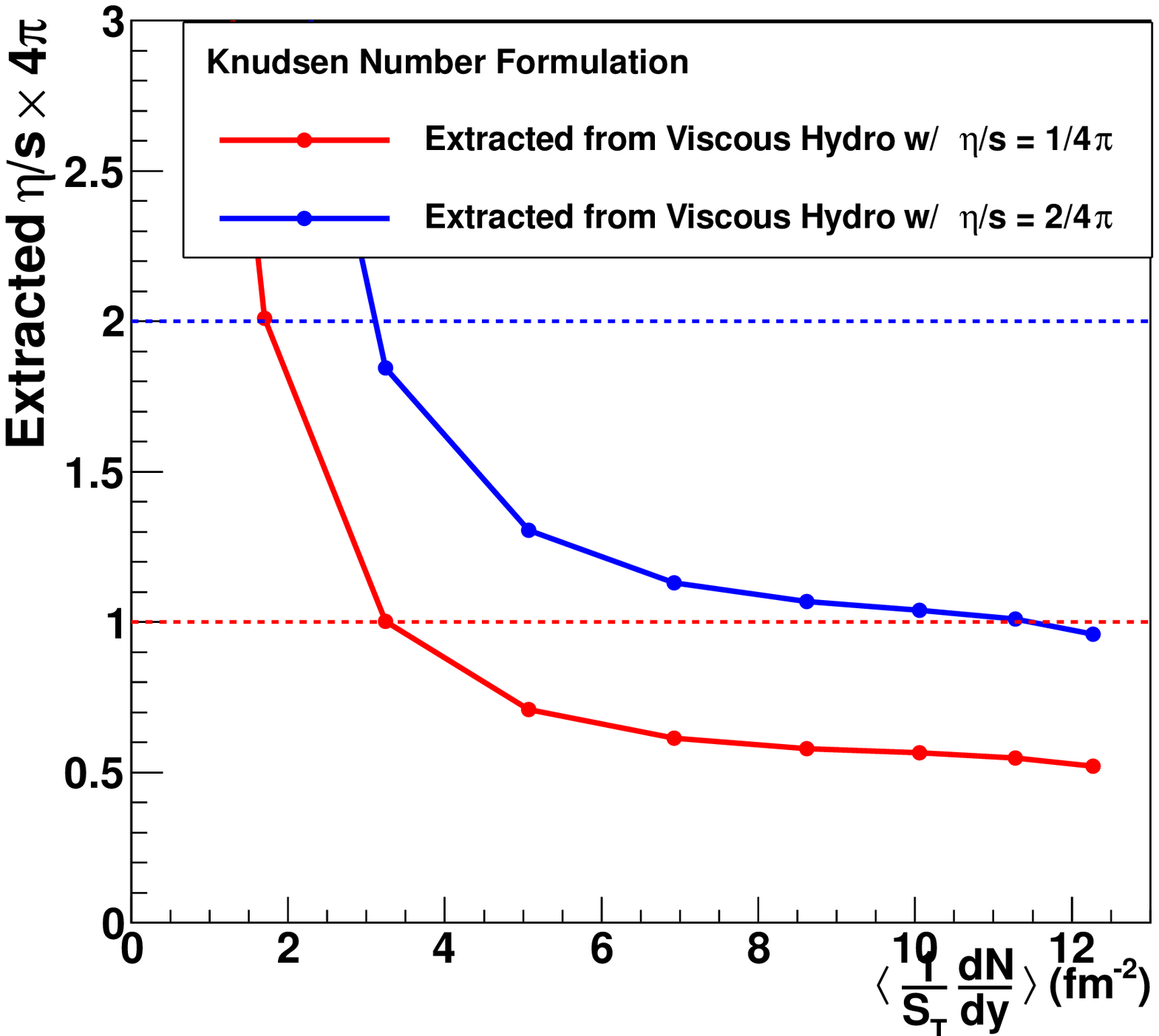}
\caption{The left panel shows results for $v_{2}/\epsilon_{part}$ as a
  function of the transverse particle density $\left< 1/S_{T} dN/dy
  \right>$ from calculations of viscous hydrodynamics with fixed
  $\eta/s$ = 0.001, $1/4\pi$, and $2/4\pi$.  The red and blue curves
  are fits to the calculation using the Knudsen number formalism.  The
right panel shows the resulting Knudsen number extraction for $\eta/s$
as a function of the same transverse particle density.  The dashed
lines represent the true $\eta/s$ for the two cases as input to the
viscous hydrodynamic model.
\label{fig_knudsen}}
\end{figure*}

It is of course possible to test this Knudsen scaling formalism by comparing
it to the results from viscous hydrodynamic models where
one knows the exact input $\eta/s$~\cite{Masui:2009pw}.   
Those authors show that a Knudsen-based scaling of the
results with $\eta/s$ exists,
but did not check the formalism for self-consistency
by using the same scaling~\cite{Drescher:2007cd} to compare  the implied value
for $\eta/s$ to the input value used in the hydrodynamic calculations.
We have done this following the exact procedure detailed in~\cite{Nagle:2009ip}
whereby one arrives at a compact relation:
\begin{equation}
{{\eta} \over {s}} = 0.32 K_{0} T \overline{R} \left[
  {{(v_{2}/\epsilon)_{ih}} \over {(v_{2}/\epsilon)}} - 1 \right]
\label{Eq:Simple}
\end{equation}
where $K_{0}$ is a constant, T is the single constant temperature,
$\overline{R} = 1/\sqrt{1/\left<x^{2}\right> + 1/\left<y^{2}\right>}$
is the characteristic scale for the strongest gradient in the initial
matter configuration, and $(v_{2}/\epsilon)_{ih}$ is the ratio
achieved in the ideal hydrodynamic limit.

Shown in Figure~\ref{fig_knudsen} (left panel) are the results from
viscous hydrodynamic calculations (using the model
from~\cite{Romatschke:2007mq}) cast in these quantities for three
different fixed (centrality independent) values of $\eta/s = 0.001,
1/4\pi, 2/4\pi$.  The two finite viscosity cases are fit using the
Knudsen number formulation, and provide a reasonable fit to the theory
points.  We note that other parameterizations described
in~\cite{Nagle:2009ip} give equally good fits.  Then utilizing the
information on the initial conditions in the model for $\overline{R}$
and assuming a temperature T = 200 MeV (as was done
in~\cite{Drescher:2007cd}), we extract the values for $\eta/s$ as a
function of centrality (shown in the right panel of
Figure~\ref{fig_knudsen}).

It it is immediately clear that the extracted values for $\eta/s$ are strongly
centrality dependent, despite the fact that $\eta/s$ was input as a
constant for all centralities. While it has been postulated that this
variation with centrality mimics viscous effects due to the finite lifetime of the
system, this  hypothesis then raises the question 
why such a large  early freeze-out effect would maintain the
expected curvature relation for the approach to ideal hydrodynamics as
a function of centrality. Even more importantly,  the numerical value for the $\eta/s$ extracted
in central A-A collisions is incorrect by a factor of two.  Taken together, 
these observations argue strongly against simply ``calibrating'' the procedure
to remove the factor of two discrepancy. Any such renormalization should
also be required to reproduce the input constancy of $\eta/s$,
which in turn would require additional centrality-dependent adjustments
to the parameters that appear in Equation~\ref{Eq:Simple}. 
Such an {\em ad hoc} procedure is clearly unsatisfactory,
especially given the availability of the very hydrodynamic calculations that
would be employed to reach a forced consistency.
The inescapable conclusion is that the most reliable method currently 
available for extraction of $\eta/s$ from experimental data is direct comparison
to the output of hydrodynamic calculations, keeping in mind the issues 
identified in the preceding sections of this paper. It should be noted that
there are additional concerns not addressed here, in particular that of 
transport in the hadronic phase. However, it is understood that the larger
mean free paths in the hadronic phase lead to larger viscosity, so that 
the correct interpretation of the values used in the hydrodynamic
calculations that ignore transport after hadronization would be upper bounds
for $\eta/s$.

\section{Summary}

In this paper, we have explored some of the various inputs regarding
hydrodynamic modeling of heavy ion collisions and the sensitivity of
the final measured $v_2$ versus $p_T$ to these inputs.  This work
has been motivated by the striking
agreement of $v_2$ measurements over an enormous range in energy.
Quantifying the precise level of agreement, and then in turn understanding the 
physics implications of the trends in the data
requires a detailed understanding of these sensitivities.  
Even at the initial stage in such a systematic exploration, the argument
that the agreement is the result of fortuitous cancellation of
effects should be viewed with great scrutiny.  As but one example,
the agreement in the $v_2(p_T)$ data between RHIC and LHC
is inconsistent with the predictions of  models~\cite{Niemi:2011ix} which 
assume significantly different transport regimes at these two
energies. 
It will be most interesting to extend
hydrodynamic calculations to compare with new data at the lower colliding
energies, although issues of baryon contributions and modeling of the
EOS will also need to be addressed. It is likely that these dynamic
effects will greatly exceed the relatively modest variations in initial 
state eccentricity, which is no more than 10\% when evaluated at the
as a function of centrality rather than number of participants.

We have also highlighted the need to take ratios which reveal that the
greatest sensitivity to viscosity effects is at lower $p_T$ where the
data is also the most accurate.    These results highlight the need
for more systematic studies and a {\em re-evaluation} of previously stated
sensitivities to the early time dynamics and properties of the medium.
Similarly, comparisons of the data to ever more sophisticated hydrodynamic and
transport calculations, rather than parameterizations 
with  underlying dynamical assumptions  unsupported by detailed examination,
are far more likely to lead to precision extraction of transport
coefficients.  
An important development in these investigations is the public availability
of modern and reliable hydrodynamic codes, which greatly leverage's
the community's ability to pursue these exciting topics.

\section{Acknowledgements}

We acknowledge useful discussions with Javier Albacete, Adrian
Dumitru, Kevin Dusling, Matthew Luzum, Michael McCumber, Jean-Yves
Ollitrault, Paul Romatschke, and Raimond Snellings.  We particularly
appreciate the public availability of these algorithms and
experimental data.
JLN acknowledges funding support from the United States Department of
Energy Division of Nuclear Physics Grant DE-FG02-00ER41152, the Discovery Center
at the Niels Bohr Institute, and the Nordea Foundation.  IGB is supported by the Danish National
Science Research Council and the Danish National Research Foundation.
WAZ is supported by U.S. Department of
Energy grant DE-FG02-86ER40281.

\end{document}